\documentclass[aps,pre,twocolumn]{revtex4}
\usepackage{graphicx}
\usepackage{amssymb}

\newcommand{\beq}{\begin{equation}}
\newcommand{\eeq}{\end{equation}}

\begin{document}

\title{Complex networks of earthquakes and aftershocks}

\author{Marco Baiesi} 
\affiliation{INFM, Dipartimento di Fisica, Universit\`a di Padova, 
I-35131 Padova, Italy.}
\affiliation{Sezione INFN, Universit\`a di Padova, I-35131 Padova, Italy.}
\author{Maya Paczuski}
\affiliation{Mathematical Physics, Department of Mathematics,
Imperial College London, London UK SW7 2AZ.}
\date{August 4, 2004}

\begin{abstract}
We invoke a metric to quantify the correlation between any two
earthquakes.  This provides a simple and straightforward alternative
to using space-time windows to detect aftershock sequences and
obviates the need to distinguish main shocks from aftershocks.
Directed networks of earthquakes are constructed by placing a link,
directed from the past to the future, between pairs of events that are
strongly correlated. Each link has a weight giving the relative
strength of correlation such that the sum over the incoming links to
any node equals unity for aftershocks, or zero if the event had no
correlated predecessors.   A correlation threshold is set to drastically
reduce the size of the data set without losing significant information.
Events can be aftershocks of many previous
events, and also generate many aftershocks.  The probability
distribution for the number of incoming and outgoing links are both
scale free, and the networks are highly clustered. The
Omori law holds for aftershock rates up to a decorrelation time that
scales with the magnitude, $m$, of the initiating shock as $t_{\rm
cutoff} \sim 10^{\beta m}$ with $\beta \simeq 3/4$.  Another scaling
law relates distances between earthquakes and their aftershocks to
the magnitude of the initiating shock.  Our results are inconsistent
with the hypothesis of finite aftershock zones.  We also find evidence that
seismicity is dominantly triggered by small earthquakes.  Our
approach, using concepts from the modern theory of complex networks,
together with a metric to estimate correlations, opens up new avenues
of research, as well as new tools to understand seismicity.
\end{abstract}

\maketitle % YOU MUST USE THE \maketitle COMMAND

\section{introduction}\label{sec:intro}

Seismicity is an exceedingly intermittent phenomena \citep{kagan94}
exhibiting strong correlations in space and time.  Since the seismic
rate increases sharply after a large earthquake in the region, events
typically are classified as aftershocks or main shocks, and the
statistics of aftershock sequences are studied.  Usually, aftershocks
are collected by counting all events within a fixed space-time window
~\citep{gardner74:_after_window,keilis80:_after_window,knopoff82:_after_window,knopoff00:_after_window}
following a main event.  The size of the window may vary or scale with
the magnitude of the main event \citep{kagan02:_after-zone}, and other
refinements have been made \citep{helms02:_trig}.  However, this
method does not allow a likelyhood to be estimated that an event
thereby collected is actually correlated to the main event under
consideration. As a result, no straightforward algorithm exists to
decide if the space-time windows are too large or too small for
minimizing errors in the procedure. Also, the method cannot be easily
extended to examine remote triggering \citep{remote93,remote01} where
main shocks may trigger aftershocks at great distance in space, or
perhaps far away in time.  In addition, aftershocks may have several
preceding events to which they are correlated, perhaps with
overlapping space-time windows. Using conventional methods the conjecture
that aftershocks can rumble on for centuries~\citep{kagan_quote} cannot
be tested.

In fact, a growing body of work indicates that the distinction
 between aftershocks and main shocks is
 relative~\citep{net_eq1,kagan_quote}. There is no unique operational
 way to distinguish between aftershocks and main
 shocks~\citep{bak02:_unified}. They are not caused by different
 relaxation mechanisms \citep{hough,helmstetter03:_bath}.  Besides, a strict distinction
  may not be the most useful way to describe the
 dynamics of seismicity.  

A particular nuisance with employing space-time windows arises from the
entanglement of a vast range of scales in space, time and magnitude in
seismicity. One scale-free
 property is the \citet{gutenberg-richter41} (G-R) distribution for
 the number of earthquakes of magnitude $m$ in a seismic region, \beq
 P(m) \sim 10^{-b\,m} \quad,
\label{eq:G-R}
\eeq with $b$ usually $\approx 1$.  A second is the fractal appearance
of earthquake
epicenters~\citep{turcotte_book,kagan94,hirata89:_d_fract}, where the
fractal dimension $d_f\approx 1.6$ in
S.~California~\citep{corral03:_unified}.  A third is the Omori law
\citep{omori94b,utsu95:_omori} for the rate of aftershocks in time,
\beq \nu(t) \sim \frac{K}{c+t}\quad,\label{eq:Omori} \eeq where $c$
and $K$ are constant in time, but depend on the magnitude
$m$~\citep{utsu95:_omori} of the earthquake.

One way forward has been suggested by \citep{bak02:_unified} who
 take the perspective of statistical physics:
  Neglecting any classification of earthquakes as main shocks,
  foreshocks or aftershocks, analyze seismicity patterns irrespective
  of tectonic features and place all events on the same footing.
  They consider spatial areas and their subdivision into square cells of
  length $L$. For each of these cells, only events above a threshold
  magnitude $m$ are included in the analysis. In this way, one can
obtain a distribution of waiting times
\citep{bak02:_unified,corral03:_unified,davidsen_goltz04} and
distances\cite{note1}
   between successive events with
  \emph{epicenters}  \emph{both
  in the same cell of linear extent} $L$.   Since both
  the threshold magnitude and the length scale of the cell (or the space window) are
  arbitrary, one looks for robust or universal features of this
  distribution that appear when these parameters are varied.

We \citep{net_eq1} have previously introduced an alternative method
 for characterizing seismicity that completely avoids
 using fixed space-time windows and, at the
 same time, makes available powerful concepts and methods that are
 being developed to describe complex networks
 \citep{albert,bornholdt02:_book,newman_siam}.  Here we extend the
 method to allow more than one incoming link per earthquake.  This
 allows the network to deviate from a tree structure and exhibit
 characteristic properties of complex networks, such as
 ``clustering''~\cite{note2}, 
which may be relevant to characterizing seismicity.
 Our method also takes into account, in an unbiased way, that
 an aftershock can be correlated to many previous events.  By
 unbiased, we mean that we do not fix any length, time or magnitude
 scales for identifying aftershocks.  Nor do we fix the number of
 events they can be aftershocks of.

\section{The Method}

A general description of our method is as follows: The first step is
to propose, as a null hypothesis \citep{jaynes_book}, that earthquakes
are uncorrelated in time. Then we detect instances when that
hypothesis is strongly violated, indicating that the opposite is true.
The second step is to assign a real number, or metric, that quantifies
the correlation between any two earthquakes, based on gross violations
of the null hypothesis.  The third step is to construct a directed
network where the events that are correlated according to the metric
are nodes connected by links.  Each link contains several variables
such as the time between the linked events, the spatial distance
between their epicenter or hypocenters, the magnitudes of the
earthquakes, and the metric or correlation between the linked pairs.
We can study the statistical properties of the network and its
ensemble of space/time/magnitude variables to gain new insights into
seismicity.  Note that many variations of the null hypothesis and
associated metric are possible, but the key feature of a useful null
hypothesis, in this context, is that earthquakes are uncorrelated in
time.  

The null hypothesis, which we  previously used, is
that earthquakes occur with a distribution of magnitudes given by the
G-R law, with epicenters located on a fractal of dimension $d_f$, at
random in time.  Of course, it is patently false that earthquakes are uncorrelated
in time.  It is also unclear if epicenters form a monofractal with dimension
$d_f \leq 2$.  The point is to look for strong violations of the null hypothesis.

Consider an earthquake $j$ in the seismic region, which
occurs at time $T_j$ at location $R_j$.  Look backward in time to the
appearance of earthquake $i$ of magnitude $m_i$ at time $T_i$, at
location $R_i$.  One can ask, how likely is event $i$ given that event
$j$ occurred where and when it did?  According to the null hypothesis, the expected number of
earthquakes of magnitude within an interval $\Delta m$ of $m_i$ that
would be expected to have occurred within the time interval
$t=T_j-T_i$ seconds, and within a distance $l=|R_i-R_j|$ meters is
 
\beq 
n_{ij} \equiv (\textit{const}) \, t\, l^{d_f}  \, 10^{-b m_i}\, \Delta m \quad.
\label{eq:n}
\eeq 
Note that the space-time domain $(t,l)$ appearing in Eq.~\ref{eq:n} is selected by 
the particular history of  seismic activity in the region and not 
preordained by any observer.  The constant term in Eq.~\ref{eq:n} is estimated
by the overall seismic rate in the region over the time span of recorded events and
is evaluated later.  However, our results are  insensitive to the 
precise value of
this constant, since its value is absorbed into a threshold we define later, $c_<$.
We find that many 
 of the statistical properties of the networks are robust with respect to
varying parameters such as $c_<$, $d_f$, and $b$. In particular, we can choose
$d_f=2$ without substantially varying the results. See also \citep{net_eq1}.

   Consider a pair of earthquakes $(i,j)$ where $n_{ij}\ll 1$; so that
the expected number of earthquakes according to the null hypothesis is
very small.  However, event $i$ actually occurred relative to $j$,
which, according to the metric, is surprising.  Hence, it is unlikely
that the pair would occur in that space-time domain if they were
uncorrelated.  A small value $n_{ij}\ll 1$ indicates that the
correlation between $j$ and $i$ is very strong, and {\it vice
versa}. By this argument, the correlation $c_{ij}$ between any two
earthquakes $i$ and $j$ can be estimated to be inversely proportional
to $n_{ij}$, or \beq c_{ij}=1/n_{ij} \quad.\eeq As we show later, the
distribution of the correlation variables $c_{ij}$ for all pairs
${i,j}$ is extremely broad. Therefore, for each earthquake $j$, some
exceptional events in its past have much stronger correlation than all
the others combined.  These strongly correlated pairs of events can be
marked as linked nodes, and the collection of linked nodes forms a
sparse network of highly clustered graphs.  Unless otherwise stated in
this work, earthquakes are linked only if their correlation value,
$c_{ij}$, is greater than $c_< = 10^4$, or the expected number of
events according to the null hypothesis, $n_{ij}$, is less than
$10^{-4}$.  The error made in ignoring weakly linked pairs of events is discussed
later.

In the language of modern complex network theory
\citep{albert,bornholdt02:_book,newman_siam}, a time-oriented weighted network
grows, where nodes (earthquakes) have internal variables (magnitude,
occurrence time, and location), and links between the nodes carry a
strength (the correlation $c_{ij}$) and are directed from the older to
the newer nodes.  Empirically, we find that both the distribution of
outgoing and incoming links are scale free.  The network is composed
of highly clustered, disconnected graphs of correlated earthquakes.
Events with incoming links, or aftershocks, typically connect to many previous
events rather than just one. However, the networks are sparse and the number of
 links in the network is much less (about 0.1\%) 
than the number of pairs of earthquakes.
We find neither that every earthquake
is correlated to every other event, nor that events typically are correlated to
zero or one previous events, but a picture in between where the number of events
an aftershock is correlated to is scale-free.

  Due to the continuous nature of
the link variable, $c_{ij}$, no event is purely an aftershock or a
main shock, and it is not possible to separate events into distinct
classes.  This is consistent with previous studies indicating 
no physical  distinction between main shocks and aftershocks
\citep{hough,bak02:_unified}.  Note that singularities in
Eq.~\ref{eq:n} are eliminated by taking a small scale cutoff in time
(here $t_{\rm min}= 60$ sec) and a minimum spatial resolution (here
$l_{\rm min} = 100$ meters).

\section{Data and parameters}\label{sec:data}
The catalog we have analyzed is maintained by the Southern California
Earthquake Data Center, and can be downloaded via the Internet at
http://www.data.scec.org/ftp/catalogs/SCSN/.  We use data ranging
from January 1, 1984 to December 31, 2003, and follow a procedure
similar to our previous work (see \citet{net_eq1}
for more details).  

\begin{table}[b]
  \caption[]
{\label{tab:sum}Network quantities for the Southern California data set, unless otherwise noted.}
\vskip4mm
\renewcommand{\arraystretch}{1.2}
%\iftwocol{\small}{}
\begin{tabular}{cll}
\hline\noalign{\vskip1mm}
Quantity & Symbol & Value                  \\
\hline
magnitude threshold & $m_<$   & 3 \\
magnitude precision & $\Delta m$   & 0.1 \\
Gutenberg-Richter exponent & $b$ & 0.95 \\
fractal dimension of epicenters & $d_f$ & 1.6 \\
fractal dimension of hypocenters & $D_f$ & 2.6 \\
number of earthquakes & $N_{\rm node}$ & 8858 \\
\noalign{\vskip1mm}\hline
\end{tabular}
\end{table}

\begin{table}[b]
  \caption[]
{\label{tab:sum2D}Parameters for the network of Southern California obtained with
the  2D version of the metric, unless otherwise noted.}
\vskip4mm
\renewcommand{\arraystretch}{1.2}
%\iftwocol{\small}{}
\begin{tabular}{cll}
\hline\noalign{\vskip1mm}
Quantity & Symbol & Value                  \\
\hline
seismicity constant (see Eq.~\ref{eq:n}) & $\textit{const}$    & $10^{-11}$\\
correlation threshold & $c_<$ &  $10^4$ \\
number of links & $N_{\rm link}$ & 166507 \\
average in-degree & $\langle k_{\rm in} \rangle$ & $18.8$ \\
number of clusters & $N_{\rm cluster}$ & $2252$ \\
\noalign{\vskip1mm}\hline
\end{tabular}

  \caption[]
{\label{tab:sum3D}Parameters for the network of Southern California obtained with
the  3D version of the metric, unless otherwise noted.}
\vskip4mm
\renewcommand{\arraystretch}{1.2}
%%\iftwocol{\small}{}
\begin{tabular}{cll}
\hline\noalign{\vskip1mm}
Quantity & Symbol & Value                  \\
\hline
seismicity constant (see Eq.~\ref{eq:n3D}) & $\textit{const'}$    & $10^{-15}$\\
correlation threshold & $c_<$ &  $10^4$ \\
number of links & $N_{\rm link}$ & $154792$ \\
average in-degree & $\langle k_{\rm in} \rangle$ & $17.5$ \\
number of clusters & $N_{\rm cluster}$ & $2327$ \\
\noalign{\vskip1mm}\hline
\end{tabular}
\end{table}

The relevant quantities for our present work are summarized in
Tables~\ref{tab:sum},~\ref{tab:sum2D}, and~\ref{tab:sum3D}.  
Events with magnitude smaller than $m_< = 3$ are discarded, and $\Delta m=0.1$. 
The number of earthquakes or nodes in the network constructed using the entire catalog is
 $N_{\rm node} = 8858$.
The $b$-value of the G-R law is $b \simeq 0.95$ for this data set,
while $d_f\simeq 1.6$ was found by ~\citet{corral03:_unified}.

We consider two closely related variants of the metric: 
in the two-dimensional (2D) version,  
the earthquake depth is not considered,
and the distance between two events $i$ and $j$ is
measured as the arc length on the Earth's surface, 
\begin{eqnarray}
l_{ij}& = R_0\arccos[& \,\sin(\theta_i)\sin(\theta_j) + \nonumber\\
 & &   \cos(\theta_i)\cos(\theta_j)\cos(\phi_i-\phi_j)\,]\quad,
\end{eqnarray}
where the Earth radius is $R_0=6.3673 \times 10^6$ meters, and ($\theta_i$,$\phi_i$)
are the latitude and longitude, in radians, of the epicenter of th $i$'th event in the catalogue.

The second version (3D) takes into account the depth $h_i$ of each event.
Hence Euclidean distances between hypocenters are calculated,
\beq
l_{ij} = \sqrt{\sum_{a=1}^3 (x_i^a-x_j^a)^2}
\eeq with
\begin{eqnarray}
x_i^1 &=& (R_0-h_i)\cos\theta_i \cos\phi_i \nonumber\\
x_i^2 &=& (R_0-h_i)\cos\theta_i \sin\phi_i \\
x_i^3 &=& (R_0-h_i)\sin\theta_i \nonumber
\end{eqnarray}
and $d_f$ in the metric is replaced by the
hypocenter fractal dimension $D_f$, which is approximated as
 $D_f=d_f+1 \simeq 2.6$ for Southern California.
Thus, the 3D metric is
\beq
n_{ij} \equiv \textit{const'} \, t\, l^{D_f} \Delta m \, 10^{-b m_i}\quad.
\label{eq:n3D}
\eeq
Most of the statistical results we find are not sensitive to the choice of the 
metric, nor to the precise values of $b$, $d_f$, or $D_f$.
For this reason, we pick as a standard metric the 2D version
(Eq.~\ref{eq:n}), and use the 3D metric only when
explicitly stated.

The constant in Eq.~\ref{eq:n} was estimated to be $\textit{const}
=10^{-11}$ for the 2D metric using the same method as in
\citet{net_eq1}.  However in that work a fractal dimension of
$d_f=1.2$ was inadvertently used to estimate $\textit{const}$,
resulting in a different value.  Similarly, here we compute
$\textit{const'} =10^{-15}$ for the 3D metric, Eq.~(\ref{eq:n3D}).
Both values give consistent results, but they are not expected to be
precise due to the high variability of seismicity rates in the region
even over a time span of years.  However, varying the constants,
$\textit{const}$ and $\textit{const'}$, in our analysis is equivalent
to varying the correlation threshold for linking events, $c_<$.  We
observe that many of the statistical results presented below are
robust to variations of $c_<$.  This is primarily for two
reasons. First, as we will show the distribution of link weights
$P(c)$ is very broad and doesn't pick out preferred values. Second,
the distributions we compute are weighted using the link
weight. Reducing the threshold for included links only adds earthquake
pairs that give progressively lower contribution to the final
correlation structure. We give later a numerical estimate for the
error made in throwing out these degrees of freedom. The advantage,
obviously, is that a sparse network with $c_<$ chosen appropriatly,
enables us to vastly reduce the size of the data set from
approximately 10-100 Gigabytes to around 10 Megabytes or so, without
losing important information about correlations between earthquakes.
The network allows a 'renormalization' which removes irrelevant
degrees of freedom, or links with low weights $c_{ij}$ while keeping
 important ones.

%FIG %%%
\begin{figure*}[!tb]
\includegraphics[angle=-90,width=83mm]{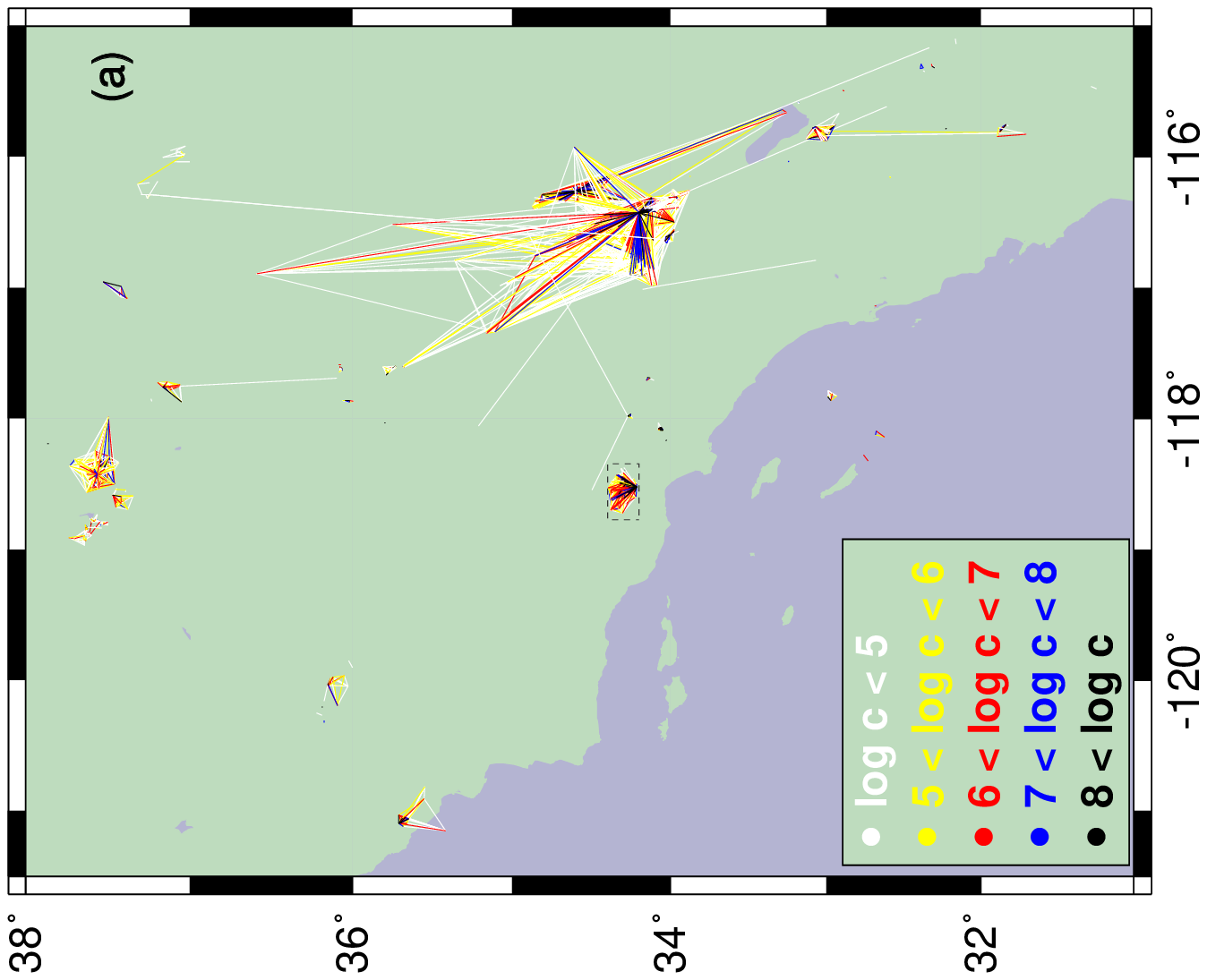}
\hskip 6mm
\includegraphics[angle=-90,width=83mm]{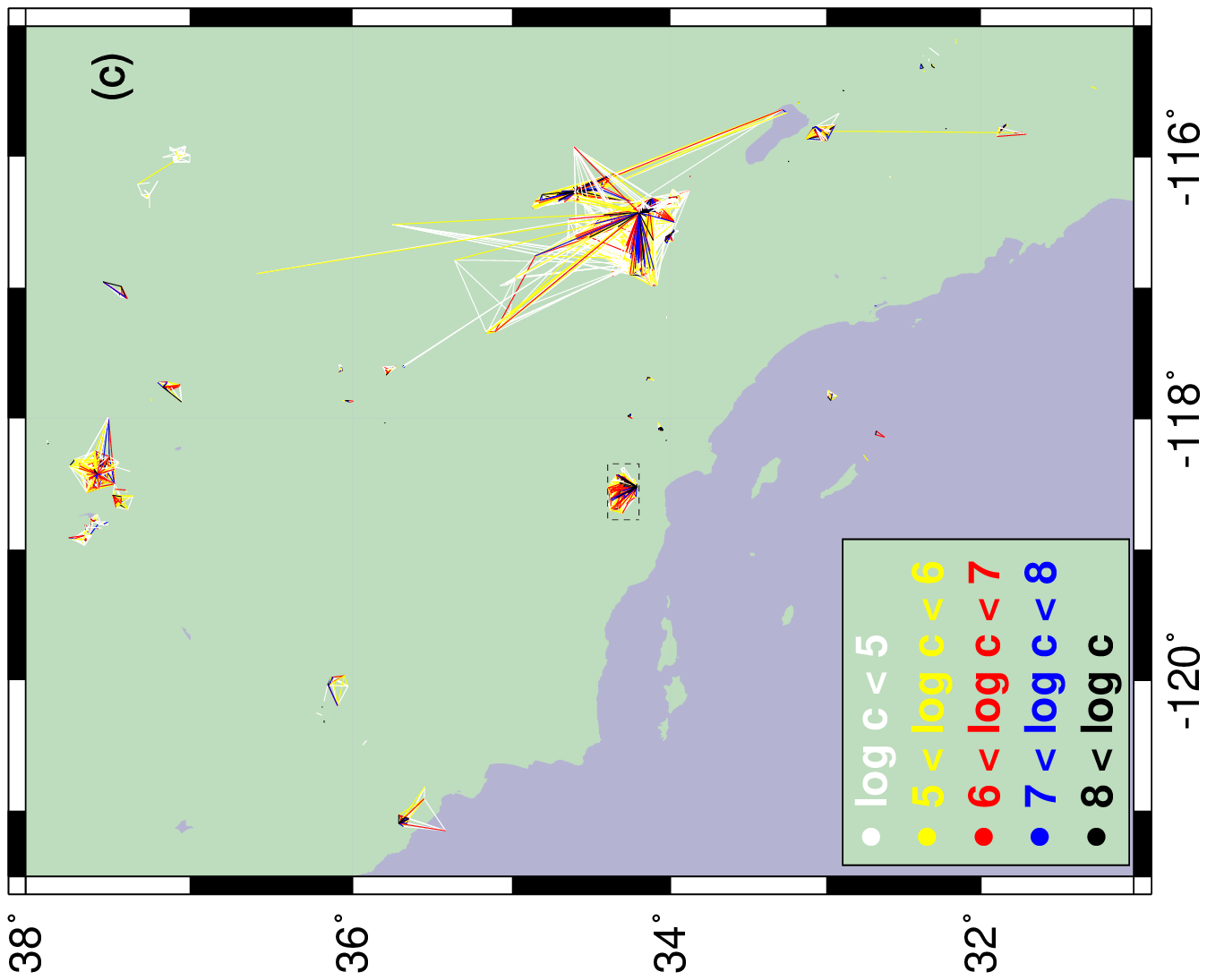}
\vskip 3mm
\includegraphics[angle=-90,width=83mm]{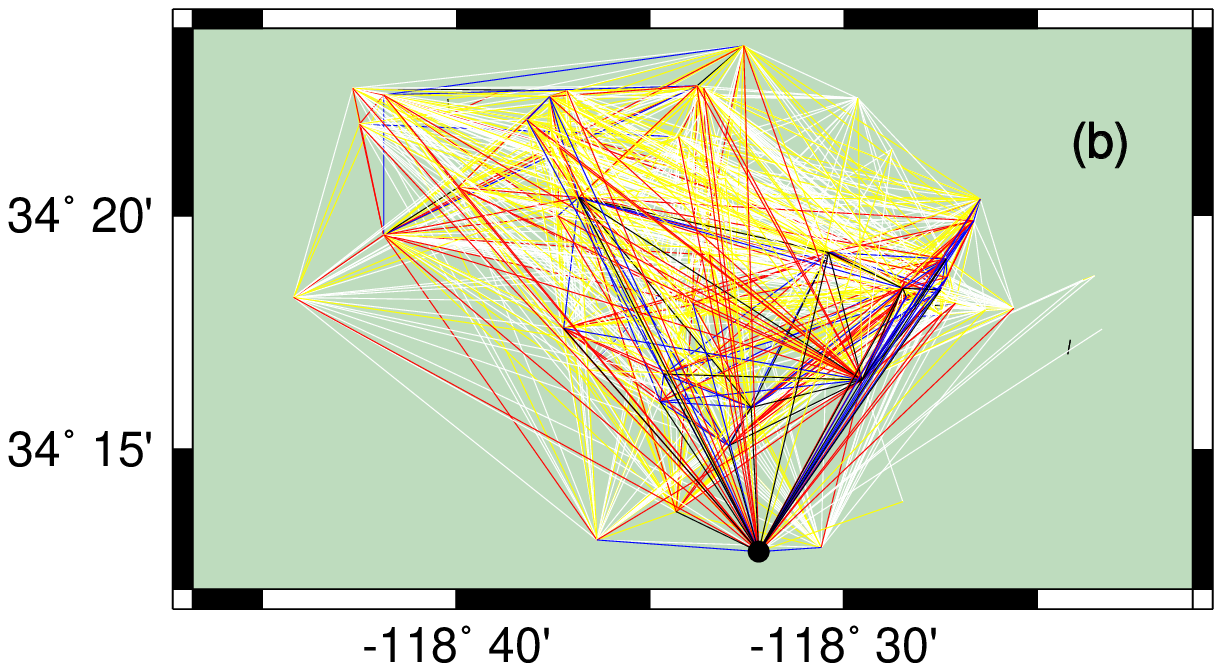}
\hskip 6mm
\includegraphics[angle=-90,width=83mm]{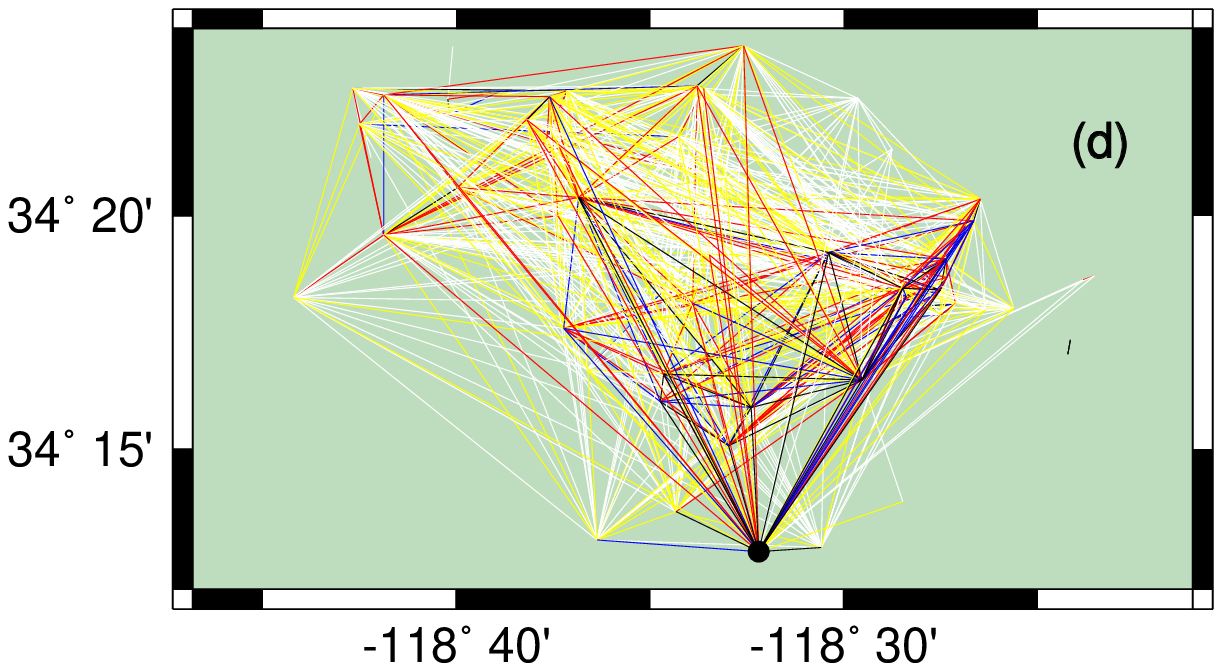}
\caption{(a) Scale free network of earthquakes obtained with the 2D
metric using $m_< = 4$ and $c_< = 10^4$. This network has $791$ nodes,
and its $7931$ links follow the color code in the legend.  Several
clusters are evident, the biggest being related to the Landers
earthquake.  The Northridge cluster, enclosed within a dashed box in
(a), is enlarged in (b), where the solid black dot represents the
epicenter of the Northridge event.  Figures (c) and (d) are obtained
using the 3D metric with $m_< = 4$ and $c_< = 5\times 10^3$ giving
$7947$ links, very close to the number of links in the 2D version.
Note that the networks found using these two  metrics are
similar, indicating that the method is robust to 
variations in the metric.
\label{fig:net}}
\end{figure*}
%FIG %%%

\section{Results}\label{sec:results}
Networks constructed using our method are shown in
Fig.~\ref{fig:net}. For comparison, networks obtained with the 2D
metric [Fig.~\ref{fig:net}(a) and (b)] and with the 3D metric
[Fig.~\ref{fig:net}(c) and (d)] are both displayed.  For visual
clarity,  a higher threshold for earthquake magnitudes $m_< = 4$ was used 
 in order to reduce the number of nodes and links in the figure.
 Adjusting the parameter $c_<$ slightly, two networks
with a similar number of links, with very similar clusters of
correlated events are formed. There is a more abundant presence of
long distance links in the 2D version but the similar details in the Northridge
clusters [Fig.~\ref{fig:net}(b) and (d)] suggest that it is mainly 
seismic history that determines the network structure, rather than
the precise details of our metric.

\subsection{Explanation of Method}\label{sec:method}

\begin{figure}[!tb]
\includegraphics[width=81mm,angle=0]{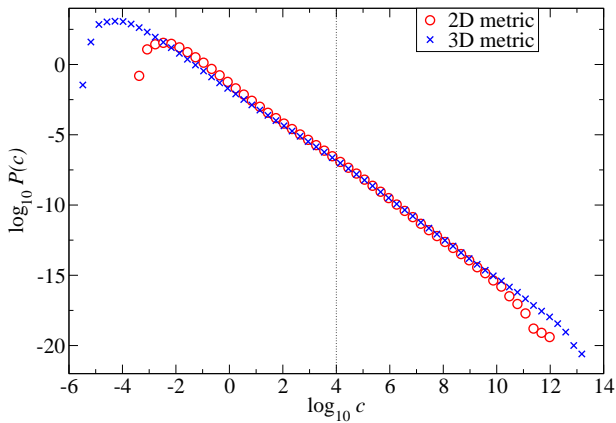}
\caption{The probability distribution of the correlation, $c$, between
all earthquake pairs in the data base, with $m_< =3$, using both the
2D metric and  the 3D one. They are scale free distributions over
many orders of magnitude.  The threshold $c_<=10^4$ where correlations
are considered significant and links are made is indicated in the
figure. Note that, with that threshold, most links are eliminated from the network,
giving a reduced data set to examine seismic properties.
\label{fig:inv_nall}}
\end{figure}

Fig.~\ref{fig:inv_nall} 
shows the probability distribution of correlation values, 
$P(c)$, obtained by
sampling the  values $c_{ij}$ over all earthquake pairs in the data set. 
 It is a fantastically broad distribution that exhibits power law
behavior over  sixteen orders of magnitude (in the 3D case):
\begin{equation}
P(c) \sim c^{-\tau} 
\end{equation}
with $\tau=1.43 \pm  0.03$ using the 2D metric and  $\tau=1.38 \pm  0.03$ using the 3D one.

%FIG %%%
%FIG %%% 

 Given such a broad distribution, for any earthquake $j$, some
extreme events $i$ exist whose correlation $c_{ij}$ are much larger
than all the others.  Therefore, it makes sense to represent these 
earthquake pairs as nodes that are linked, while not linking pairs
that have much smaller values of $c_{ij}$.  Then the sequence of
earthquakes may be usefully represented as a sparse network, where
links exist between the most strongly correlated events, i.e.\ those pairs $(i,j)$
where $c_{ij} > c_<$.  Hence a natural decomposition of the network into disconnected clusters
is achieved, where the first earthquake in the directed cluster has no incoming link, or
correlation variable into it greater than $c_<$. Clearly, the first earthquake in the entire
catalogue also  no incoming link.
The correlated events are
reliably detected when $c_<$ is greater  than one but not extremely large.  In the latter case,
correlated events detach, and a very fragmented network appears.
For small  $c_<$ some uncorrelated events make links, and a
giant cluster appears.  
Both for the 2D and for the 3D case we set $c_< =10^4$, unless otherwise noted, obtaining a similar number of links
in the realization of the networks.

\subsection{The Scale-Free Network}
\begin{figure}[!tb]

\includegraphics[width=81mm,angle=0]{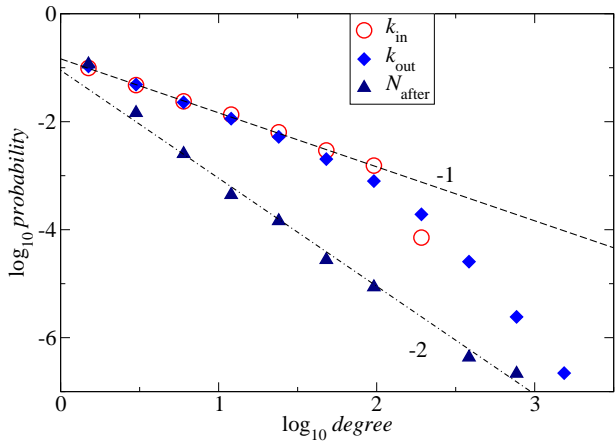}

\caption{The in-degree and out-degree distributions of the network of earthquakes and
aftershocks.  The out-degree, $k_{\rm out}$, is the number of outgoing links
from an earthquake, linking it to its aftershocks. The in-degree, $k_{\rm in}$, is the number of incoming links
to an earthquake, linking it to its main shocks.
  These two distributions are  similar.  Also shown is the distribution of
the weighted number of aftershocks, $P(N_{{\rm after}})$, from any event, using $\eta=1$ in
Eqs.~\ref{eq:eta} and ~\ref{eq:kappa}. This has
the same scaling behavior as the extremal network, with $\eta \rightarrow \infty$.
\label{fig:P_k}}
\end{figure}

The resulting network of earthquakes is scale free.  As shown in Fig.~\ref{fig:P_k},
both the distribution of the number of incoming links, or the ``in-degree'' $k_{\rm in}$, to a node and the distribution
of the number of outgoing links, or the ``out-degree'' $k_{\rm out}$, to any node exhibit power law behavior, 
\beq P(k_{\rm in}) \sim 1/k_{\rm in} \,,\quad P(k_{\rm out}) \sim 1/k_{\rm out} \eeq
up to a degree $\approx 100$.

%FIG %%%
%FIG %%%

\subsubsection{Aftershocks with more than one main shock}

Since an earthquake can have more than one incoming link, in
attributing aftershocks to an event we must be careful not to
overweight aftershocks with many incoming links.  To prevent the
overcounting of aftershocks, one can consider a new event with two
incoming links, for example, to be ``half an aftershock'' of both of
its precursors, or they can be weighted in a different fashion
according to their correlation values.  In general, we can attribute
the relative correlation to previous events, so that each event
contributes a total weight of unity to the global aftershock number if
it is linked to at least one previous event, and zero otherwise, as
follows:

For each event $j$ that has at least one incoming link, so that it can
be called an aftershock, define a weight for each "parent" earthquake
$i$ it is linked to as \beq w_{ij}= {c_{ij}^{\eta} \over \sum_{k}^{\rm
in} c_{kj}^{\eta}} \quad , \label{eq:eta}\eeq where the sum is over
{\it all earthquakes $k$ with links going into $j$}.  The weighted
number of aftershocks attributable any event $i$ is then \beq N_{{\rm
after},i} = \sum_j^{\rm out} w_{ij} \label{eq:kappa} \quad .\eeq Here,
the sum is { \it over all of the outgoing links from event $i$.}  In
the limit that $\eta \rightarrow \infty$, the extremal network studied
by \citet{net_eq1} is recovered, since only the single incoming link
to aftershock $j$, with the largest correlation $c_{ij}$,
contributes. In that case, for each node, the quantity $N_{{\rm
after}}$ discussed here coincides with the quantity $k_{\rm out}$ in
\citet{net_eq1}.  In the following, we consider the case $\eta=1$.

Sampling over all earthquakes, we get a probability
distribution for the number of (weighted) aftershocks as also shown in
Fig.~\ref{fig:P_k}.  It is a power law distribution, $P(N_{\rm after})
\sim (N_{\rm after})^{-\gamma}$ scaling over more than three decades,
with an index $\gamma = 2.0(1)$.  This distribution is very close to
the distribution we obtained previously for the number of aftershocks
in the extremal network, corresponding to the limit $\eta \rightarrow \infty$.
The distribution for the number of
weighted aftershocks appears to be universal, in the sense that the power law exponent
does not depend on $\eta$, for $\eta \geq 1$. Note that chosing a positive $\eta$ gives
more weight to more strongly correlated pairs and is therefore consistent with using
a threshold $c_<$ to eliminate weakly correlated ones.

%FIG %%%
\begin{figure}[!tb]

\includegraphics[width=81mm,angle=0]{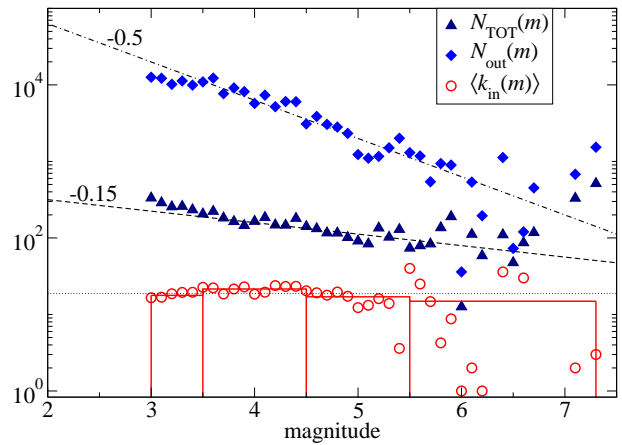}

\caption{
The total number of outgoing links from all events of magnitude $m$ (diamonds),
the total number of weighted aftershocks of all events of magnitude $m$ (triangles) and
average number of incoming links ($\langle k_{\rm in} (m) \rangle$, circles and histogram) as a function of $m$.
The first two quantities have a scaling consistent with the law $\sim 10^{\alpha m}$,
with $\alpha' \approx b - 0.5 \approx 0.45$ for outgoing links, and
$\alpha \approx b - 0.15 \approx 0.8$ for weighted aftershocks.
For $\langle k_{\rm in} (m) \rangle$
there is no evident departure from the average value $\langle k_{\rm in} \rangle \simeq 18.8$ 
(dotted line). \label{fig:alpha}}
\end{figure}
%FIG %%%

\subsubsection{Is Seismicity driven by small or large earthquakes?}
The average number of aftershocks of an earthquake of magnitude $m$
has been proposed to scale with $m$
\citep{utsu69:_alpha,kaganknopoff87:_alpha} as \beq N_{\rm after}(m)
\sim 10^{\alpha m} \label{eq:alpha} \quad . \eeq Larger main shocks
release more energy and therefore ``trigger'' more aftershocks than
smaller earthquakes.  However, smaller earthquakes are more frequent,
as indicated by the G-R relation.  The total number of aftershocks
generated by all earthquakes of magnitude $m$ is therefore given by
the product of these two relations as \beq N_{\rm TOT}(m)= N_{\rm
after}(m)P(m)\sim 10^{(\alpha -b)m} \label{eq:total} \quad . \eeq If
the exponent $\alpha >b$ then small earthquakes are the dominant
triggering mechanism for seismicity, whereas if $\alpha < b$ the large
earthquakes dominate aftershock production.  Often it is assumed that
$\alpha=b$ \citep{kaganknopoff87:_alpha,reasenberg_jones}.  

Recently,
\citet{helms02:_trig} analyzed earthquake catalogues by means of a
"stacking" method using space-time windows, and found that aftershocks
were predominantly triggered by small earthquakes. She determined the value
of the exponent $\alpha$ to be between $\alpha=0.72$ and
$\alpha=0.82$, depending on the parameters of the aftershock detection
algorithm that she used.

Fig.~\ref{fig:alpha} shows $N_{\rm TOT}(m)$ obtained using the 2D metric
with $\eta=1$.  The results shown in this figure also suggests that small
earthquakes are the dominant  mechanism driving aftershock production.  We
determine the value of the exponent $\alpha\approx 0.8$, consistent with
Helmstetter's previous findings.  Thus at this rather detailed level,
results obtained using our method are consistent with results obtained
using traditional methods of aftershock detection.  This is true
despite the fact that aftershocks in our algorithm are typically
attached to many previous events rather than just one, and no
space-time scales are used by us for aftershock identification.

Fig.~\ref{fig:alpha} also shows the total number of links emanating
from events of magnitude $m$, $N_{\rm out}(m)$.  This corresponds to
an unweighted aftershock number, with $N_{\rm out}(m) \sim
10^{-(\alpha' -b)}$ and $\alpha'=0.45$. Since both $\alpha$ and
$\alpha'$ are less than $b$, our results suggest that irrespective of
the manner in which aftershocks are weighted, small earthquakes are
the dominant mechanism driving aftershock production.  Note however, that the
 largest events may appear to present a deviation from this behavior, but the statistical
uncertainties of single events are large. 

\subsubsection{B{\aa}th's Law}

In Fig.~\ref{fig:alpha}, we also show the dependence of the number of
incoming links to a node on its magnitude.  The quantity $\langle
k_{\rm in}(m)\rangle$, is the average number of incoming links to
earthquakes of magnitude $m$.  This quantity is independent of
earthquake magnitude for $3 \leq m \leq 5$.
For larger magnitudes, the poor statistics forces us to average $k_{\rm in}(m)$ over wider bins,
chosen so that there is a significant number of events inside each one.
These averages are indicated as horizontal lines  in Fig.~\ref{fig:alpha},
while vertical lines denote the bin boundaries.
The averages do not show any detectable trend for larger earthquakes.  

Our results suggest that earthquakes of all magnitudes are equally
likely to be aftershocks, and support the conclusion reached by
\citet{helmstetter03:_bath} that observations of B{\aa}th's law are
 due to biases in labelling earthquakes as aftershocks.  According
to B{\aa}th's law \citep{bath_law}, the average magnitude difference
between a main shock and its largest aftershock is around 1.2,
independently of the main shock magnitude.  Of course, the definitions
we use here for main shocks, as nodes giving outgoing links, and
aftershocks, where nodes have incoming links (so that a single event
can be both a main shock and aftershock), differs from the standard
definition.

\subsubsection{Clustering of nodes}
Among the concepts in network theory that may be useful to characterize 
seismicity, and are not accessible via other approaches, the clustering of nodes deserves particular attention.
Indeed, the clustering in space and time of earthquakes can be quantified in their
network by the clustering coefficient.
The clustering coefficient of a node $i$ is the number $\Delta_i$ of linked triangles  
it forms with its $k_i$ neighbors (or equivalently, the pairs of linked neighbors) 
divided by the maximum number of linked triangles
it could potentially have ($k_i(k_i-1)/2$), i.e.,
\beq C_i =  \frac{2 \Delta_i}{k_i(k_i-1)} \label{eq:Ci}\quad.\eeq
This definition ignores the directionality of links.  Thus, in this formula
the degree of node $i$,  is the sum of its incoming and outgoing degrees,
i.e. $k_i=k_{i, \rm in} + k_{i, \rm out}$.
In all cases $0\le C_i\le 1$, and $C_i=0$ if less than two links 
are joined to node $i$, or if no links between its linked neighbors are present,
while $C_i=1$ only if all neighbors are linked to each other.

Using Eq.~(\ref{eq:Ci}) to compute the average clustering coefficient of the network,
\beq C = \frac{1}{N_{\rm node}}\sum_{i=1}^{N_{\rm node}} C_i\label{eq:C}\quad
,\eeq
 we obtain $C=0.50$ for $m_<=3$.   This value is relatively stable with respect to
variations of $m_<$. For instance, with  $m_< = 4.5$ we get $C=0.55$. The same values are obtained
for the 3D version of the metric. These are remarkably high values of $C$,  
compared to many other complex networks, such as technological or biological ones~\citep{newman_siam}.

%FIG %%%
\begin{figure}[!tb]
\includegraphics[width=81mm,angle=0]{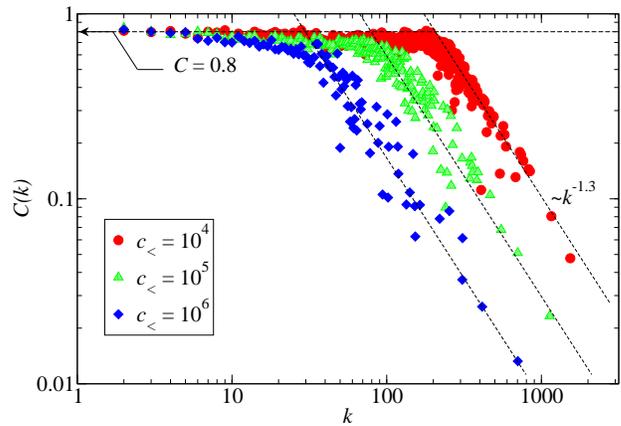}
\caption{The network density, or clustering coefficient as a function of the degree, $k$, of a node,
where $k=k_{in} + k_{out}$.
For small values of $k$, the clustering coefficient
is independent of $k$ and $c_<$, and   $C\approx 0.80$. For large values
of $k$, $C(k)$ tends toward power law behavior $C(k) \sim k^{-\delta}$, with $\delta
\approx 1.3$. The power law regime takes place at smaller $k$ for larger thresholds
$c_<$. \label{fig:C}}
\end{figure}
%FIG %%%

\paragraph{Universal clustering properties of seismic networks}

The average of the clustering coefficient 
can be performed over   nodes with the same degree $k$.
This quantity is shown in Fig.~\ref{fig:C}, where one can observe that it does not
depend on $k$ for small values of $k$ and approaches a power law
$C(k) \sim k^{-\delta}$ with $\delta \approx 1.3$ for large values of $k$.
This power law behavior is typically  found in networks with a modular 
structure~\citep{ravasz03:_hier}.
At small $k$, $C(k)$ approaches a universal value approximately equal to
$0.8$, which is independent of $k$ and of the thresholds $c_<$ and $m_<$ used
to construct the network.
The power law exponent $\delta$ at large values of $k$ also
appears to be independent of $c_<$, and may also  be a
universal quantity for seismic networks.

\subsection{Scaling Law for Aftershock Distances}

In the network constructed using the 2D metric,
the link length, $l$, is the distance between the epicenters
 of an earthquake and one of its aftershocks, weighted according to
the link weight $w$.  In the corresponding network constructed using the
3D metric, the link length is the distance between the hypocenters of an
earthquake and one of its aftershocks, weighted according to the link weight
$w$.
The distribution $P_m(l)$ of
 link lengths depends on the magnitude $m$ of the predecessor, being
 on average greater for larger $m$.
 To compute this distribution,  
we put the weight of each link into a bin corresponding to its $l$ value
and the magnitude of the predecessor $m$ to get $P_m(l)$.
A maximum in the distribution occurs, which
 shifts to larger $l$ on increasing $m$, as shown in
 Fig.~\ref{fig:linkl}.  This behavior is superficially
  consistent with using larger
 space-time windows to collect aftershocks from larger events, or the
 \citet{kagan02:_after-zone} hypothesis of aftershock zone scaling with main
shock magnitude.

\subsubsection{Comparison with Aftershock Zone Scaling}
 
 It is widely believed that an aftershock zone exists which is
 equivalent to the rupture length. Within the aftershock zone,
 earthquakes generate aftershocks, while outside the zone they do not. The
 rupture length, $R$, is believed to scale as $R \sim 10^{0.5m}$ with
 the magnitude of the main shock.  This is a restatement of
 the relation derived by \citet{kanamori-anderson75}, who argued
that the seismic moment $M\sim 10^{1.5 m}$
 scales with $R$ as $M\sim R^3$, at least for intermediate magnitude
 earthquakes. For a generalization to all earthquakes, see
 \citet{kagan02:_after-zone}.  In this scenario main shocks of all
 magnitudes generate aftershocks at the same rate within their
 respective aftershock zones, so that the greater number of
 aftershocks coming from large events is due solely to their larger
 aftershock zones.  Needless to say, the observation of aftershock
 zone scaling is based on the idea that the aftershock zone is finite
 -- on the order of tens of kilometers for large main shocks.

%FIG %%%
\begin{figure}[!tb]
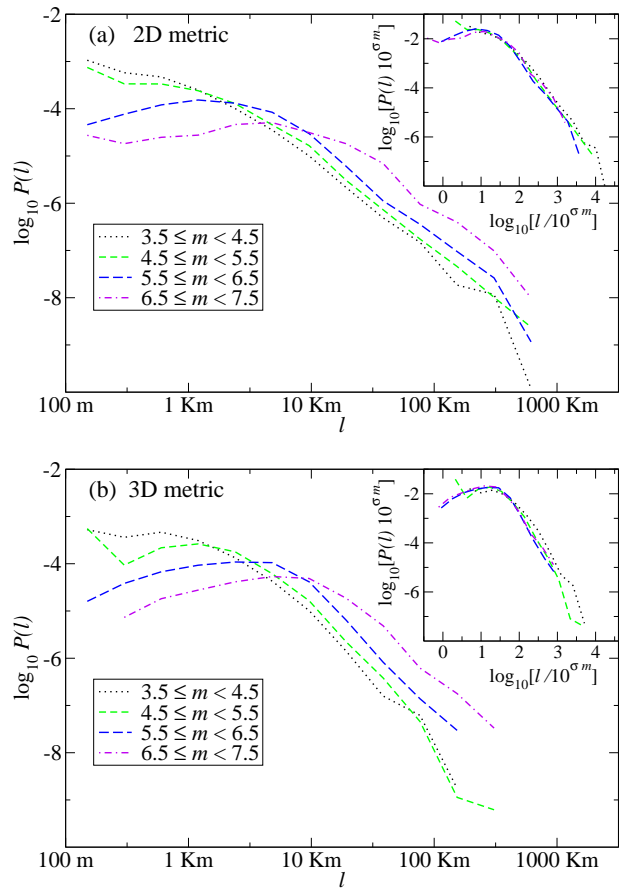

\includegraphics[width=81mm,angle=0]{ms-baiesi-f06a.eps}
\vskip 3mm
\includegraphics[width=81mm,angle=0]{ms-baiesi-f06b.eps}
\caption{Link length distribution for different magnitudes of the
emitting earthquake, (a) for the 2D case and (b) for the 3D one.  
The length where the maximum in the distribution occurs increases with
magnitude roughly as $l_{\rm max}\sim 10^{0.37 m}$ in both cases.  Both distributions
also have a fat tail, extending up to hundreds of kilometers
even for intermediate magnitude events.  These distributions are
consistent with a hierarchical organization of events, where big
earthquakes preferentially link at long distance with intermediate
ones, which in turn link to more localized aftershocks, and so on.
Insets: Distributions rescaled according to Eq.~\ref{eq:Pl_resc} with
$\sigma=0.37$ and with $m$ equal to the central magnitude of the range
for each distribution.
\label{fig:linkl}}
\end{figure}
%FIG %%%

 In contrast, we find the distribution of lengths between main shocks and
 their aftershocks exhibits  {\it{no cutoff at large
 distances,}} but rather decays slowly as a power law with $l$, up to
 the linear extent of the seismic region covered by the catalog, hundreds of
kilometers.  The
 two distributions are both consistent with a scaling ansatz: \beq
 P_m(l) \simeq 10^{-\sigma m} F \big( l/10^{\sigma m} \big) \,
\label{eq:Pl_resc}
\eeq where $l$ is measured in meters and $F(x)$ is a scaling function.
Remarkably in both cases, $\sigma\approx 0.37$.  Note in particular that
$\sigma \neq 0.5$.

For $x\gg 1$, the tail of the scaling function is a power law,
i.e. $F(x)\sim x^{-\lambda}$ with $\lambda \approx 2$ (2D) or
$\lambda\approx 2.6$ (3D).  The results obtained using the data
collapse technique applied using ansatz (\ref{eq:Pl_resc}) are shown
in the insets of Fig.~\ref{fig:linkl}(a) and (b).  Such slow decays at
large distances calls into question the use of sharply defined space
windows for collecting aftershocks, as already pointed out 
by~\cite{ogata98:_space}.  The length scale we find $l_m^*\sim 10^{0.37 m}$
to describe the {\it fat-tailed distribution of distances between
earthquakes and their aftershocks} should not be confused with the
{\it scaling of a finite aftershock zone} as proposed by
\citet{kagan02:_after-zone}.  Instead, our results are consistent with
observations of remote triggering of aftershocks by~\citet{remote93}
and~\citet{remote01} as well as the observation that the distribution
of distances between subsequent earthquakes in regions of size
$L$ is a power law, not trivially given by the correlation
dimension, $d_2$ of earthquakes, and which is cutoff only by the size of the
region $L$~\cite{note1}.

\subsection{The Omori Law for Earthquakes of All Magnitudes}

\begin{figure}[!tbp]
\includegraphics[width=81mm,angle=0]{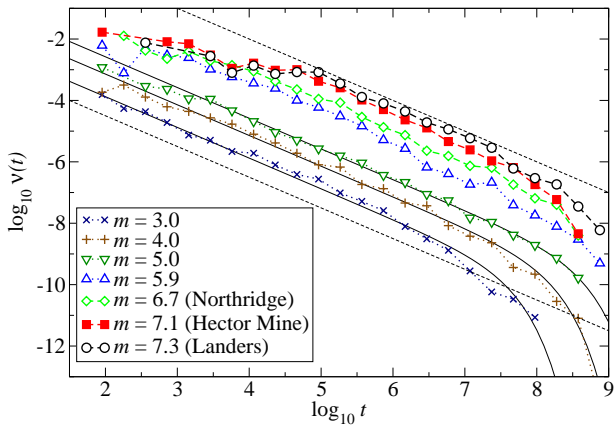}
\caption{The Omori law for aftershock rates. These rates are measured for 
aftershocks linked to earthquakes of different magnitudes.  
For each magnitude, the rate is consistent with the original
Omori law, Eq.~\ref{eq:Omori}, up to a cutoff time that depends on
$m$.  As guides to the eye, dashed lines represent a decay $\sim
1/t$. The dense curves represent the fits obtained by means of
Eq.~\ref{eq:fit} for $m=3$, $m=4$, and $m=5$.
\label{fig:Omori}}
\end{figure}

Figure~\ref{fig:Omori} shows the rate of aftershocks for the Landers,
Hector Mine, and Northridge events, obtained with the 2D metric.  
The weights, $w$, of the links to aftershocks occurring at time $t$
after one of these events are binned into geometrically increasing
time intervals.  The number of weighted aftershocks in each bin is then divided
by the temporal width of the bin to obtain a rate of weighted aftershocks
per second.  The same procedure is applied to each remaining event,
not aftershocks of these three.  An average is made for the rate of
aftershocks linked to events having a magnitude within an interval $\Delta m$
of $m$.  Figure~\ref{fig:Omori} also shows the averaged results for $m=3$
(1871 events), $m=4$ (175 events), $m=5$ (28 events) and $m=5.9$ (4
events).

%FIG %%%
%FIG %%%
%FIG %%%
\begin{figure}[!tbp]
\includegraphics[width=81mm,angle=0]{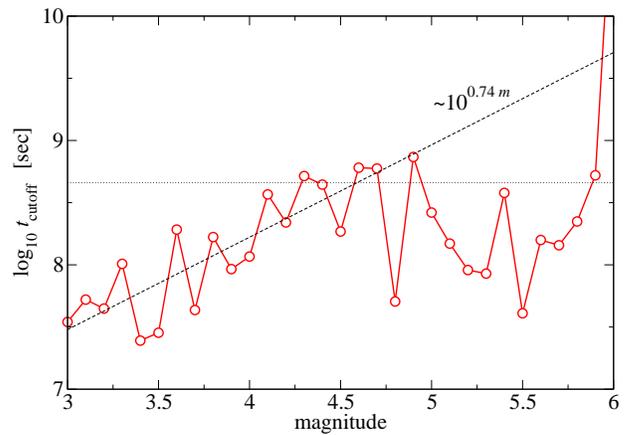}
\caption{Decorrelation time, $t_{\rm cutoff}$ of the aftershock rates to fall out of
the Omori regime, as a function of the earthquake magnitude.
The horizontal line indicates 20 years, i.e. the time span of the catalogue.
The dashed line is the interpolation given in Eq.~\ref{eq:t_cutoff}. 
\label{fig:t_cutoff}}
\end{figure}
%FIG %%%

The collection of aftershocks linked to earthquakes of all magnitudes
is one of the main results of our method. Even intermediate magnitude
events can have aftershocks that persist up to years.  Earthquakes of
all magnitudes have aftershocks which decay according to the Omori law
\citep{omori94b,utsu95:_omori}, \beq \nu(t) \sim
\frac{K}{c+t}\quad,\label{eq:Omori2} \quad {\rm for}\ t<t_{\rm cutoff}\eeq
where $c$ and $K$ are constant in time, but depend on the magnitude
$m$~\citep{utsu95:_omori} of the earthquake.  We find that the
Omori law persists up to a decorrelation time $t_{\rm cutoff}$ that also depends on
$m$.

A rough  estimate of the decorrelation times can be extracted by non-linear fits of 
$\log_{10}\nu_m(t)$ vs $\log_{10} t$, using an interpolating function
\beq
\nu_m(t)\sim t^{-1}e^{-t/t_{\rm cutoff}}\quad.\label{eq:fit}\eeq
The range of the fit excludes short times, where the
the aftershock rates are not yet scaling as $1/t$.  The short time deviation from power law behavior is
 presumably due to saturation of the detection system,
which is unable to reliably detect events happening at a fast rate.  However, this problem does not
occur at later times, where the rates are lower.
Some examples of these fits are also shown
in Fig.~\ref{fig:Omori} for the intermediate magnitude events.

In Fig.~\ref{fig:t_cutoff} we show the resulting values of $t_{\rm cutoff}$, for $m$ up
to $6$. The horizontal dotted line represents the time span of the catalogue we study, which
precludes accurate estimates  of much longer $t_{\rm cutoff}$.
Thus, from the rates of aftershocks of events with 
 $3\le m \le 4.6$, where the time span of the catalogue is comparable or
longer than the estimated decorrelation time, we find that
the increase of $t_{\rm cutoff}$ with $m$ can be fitted by the function
\beq
t_{\rm cutoff}(m) \simeq 10^{5.25+0.74 m}\quad {\rm sec} \quad ,\label{eq:t_cutoff}
\eeq
represented as a dashed line in Fig.~\ref{fig:t_cutoff}.
It roughly corresponds to $t_{\rm cutoff}\approx 11$ months for $m=3$,
and to $t_{\rm cutoff} \approx 5$ years for $m=4$.
An extrapolation yields $t_{\rm cutoff} \approx 1400$ years for an event
with $m=7.3$ such as the Landers event!
However, we stress that Eq.~\ref{eq:t_cutoff} is just rough
estimate of $t_{\rm cutoff}(m)$.

Note that \citet{helms02:_trig} also found an Omori law for
aftershocks of earthquakes of all magnitudes using finite space-time windows.
However, for this reason, she was not able to estimate the decorrelation of aftershocks
 or the cutoff  in the
duration of the Omori regime for different magnitudes.

\section{Discussion}

At present, we are unaware of any reliable method to determine the
best metric.  Thus, the best route  to study how sensitive
the results are to variations of the metric or to the parameters of the metric.
Although we have not yet made an exhaustive and detailed study to determine
which properties may be universal and hold for many different metrics, several
general conclusions are already apparent.

\subsection{Robustness with respect to changes in parameters}
Many of the statistical results we find are relatively robust with
respect to variations in the metric.  For instance, using both the 2D
metric (Eq.~\ref{eq:n}) and the 3D metric (Eq.~\ref{eq:n3D}), similar
networks are found as indicated qualitatively in Fig.~1.  In addition,
the scaling behaviors demonstrated in Figs. 2-7 are independent of the
metric, with the notable exception of the exponent $\lambda$
characterizing the fat tailed distribution of aftershock distances.
However, the exponent $\sigma$ for the rescaled variable combining
main shock magnitude and aftershock distance is independent of the
metric, as is the Omori behavior. Furthermore, the distribution of
correlations $P(c)$ depends only weakly on the metric, and the
scale-free and clustering properties of the network are insensitive as
well.

One could object that the values of $b$, $d_f$ and $D_f$ can depend on
the region of the Earth being considered, or may fluctuate depending
on the specific fault zone being studied.  However, the statistical
results we find, as shown in the Figures, are also robust to
variations in either of these parameters, or of the threshold $m_<$.
This robustness was also found in the
\citet{net_eq1} studies of the extremal earthquake
network.  For instance,
varying $d_f$ over a wide range, from $1$ to $2$ does not alter
considerably the distribution of incoming or outgoing links.  The
distribution of correlations $P(c)$, is even more insensitive to
variations of $b$ and $d_f$.  Also the Omori law with $p\approx 1$,
shown in Fig.~\ref{fig:Omori}, does not depend sensibly on the
parameters, and holds for aftershocks linked to earthquakes of all
magnitudes.

Our interpretation of this observed robustness is that the correlation
structure of seismicity is unambiguous and clear-cut, and has a network
structure similar to other complex networks.
Even if we use an approximate measure, or metric, the underlying
correlations are sufficiently strong that they survive the approximation
and can be reliably detected.

\subsection{Errors and Data Set Reduction}

One could also object that the parameter $c_<$ is arbitrary, and its choice
plays a similar role to
 choosing space-time windows in the traditional manner.  However,
one can consider all pairs of earthquakes, using their weights $c_{ij}$, so the
parameter $c_<$ is conceptually unnecessary. This differs from
the {\it necessity} of choosing space-time windows in the traditional approach.
However, as a practical matter, to reduce the size of the data set, it is useful
  to choose a particular $c_<$, and thereby construct
a sparse network.  The choice involves a trade-off
between the amount of data stored, and the accuracy of the representation of
seismicity one can make using that data set.  From the distribution $P(c)$ shown
in Fig.~2, and from the average number of incoming links $\langle k_{in}\rangle$ with
a given choice of $c_<$,
we can estimate the error made in throwing out  weak links. 
The average correlation
$c$ contribution from all of
 the $\approx N_{\rm node}=8858$ incoming links that are pruned from any earthquake
when imposing the threshold $c_<$ is
\begin{equation}
N_{\rm node} \int_{c_{min}}^{c_<} c P(c)dc \simeq A N_{\rm node} c_<^{2-\tau} \quad , 
\end{equation}
where $A$ is a constant given by the amplitude of $P(c)$, and $c_{min}$ is the minimum
value of $c$ observed in the measurement of $P(c)$.

The average correlation contribution from the incoming links actually represented
in the network with that choice of $c_{<}$ is
\begin{equation}
\langle k_{in}\rangle\int_{c_<}^{c_{max}} c P(c) dc 
\simeq A \langle k_{in}\rangle c_{max}^{2-\tau} \quad , 
\end{equation}
with $c_{max} \approx 10^{12}$ for the 2D metric. The relative error with a given
$c_< $ can thus be estimated as
\begin{equation}
{\rm Error}= 
\frac{N_{\rm node}}{\langle k_{in}\rangle }\big(\frac{c_<}{c_{max}}\big)^{2-\tau}
\quad .
\end{equation}
With our choice $c_<=10^4$ we get an estimate of the relative error to be approximately
one percent, or ${\rm Error} = 0.013$ with the fraction the data set stored 
$N_{links}/N_{\rm node}^2 \approx 0.002$. In other words, throwing out about
$99.8\%$ of the data
set we can accurately represent the correlation structure of seismicity using
a sparse network with an estimated error of order one percent.  Conversely,
we are not aware of any quantitative estimate of the error with particular choices
of space time windows.

\subsection{Bench mark test for models of seismicity}

The statistical properties of the network of seismicity we find can be
used to test various models of seismicity. A self-organized critical
model proposed by \citet{ofc} exhibits a universal Gutenberg-Richter
law for earthquakes \citep{lise}, independent of the dissipation
parameter, as well as foreshocks and aftershocks
\citep{hergarten}. However, no evident self-organized spatial
structure corresponding to the recurrence of earthquakes on a
heterogeneous system of faults exists. For this reason, we believe it is unlikely
that this model can reproduce the observed network properties of
seismicity. Although no satisfactory dynamical model of the
self-organization of the Earth's crust and resultant seismicity exists
at present, a stochastic branching process, known as the ETAS model
\citep{kaganknopoff87:_alpha,ogata98:_ETAS}, or its spatially
extended version \citep{ETAS:se} could be tested by constructing a
network using our method for particular realizations of that process
in space, time and magnitude, and comparing
with our results. For instance, the appearance of the scaling variable
$l10^{-\sigma m}$ combining spatial
distances with main shock magnitude could be ascertained.
  Since the distance variable between mother daughter pairs in the
spatially extended ETAS model is chosen from
a power law distribution, $\Phi(\vec r)$,
independent of the
 parent's magnitude, this model is unlikely to reproduce observed behavior and
would have to be modified. Conversely, one could also check if our method
of constructing networks linking main shocks and aftershocks correctly identifies
mother-daughter pairs given by the
 algorithm of the ETAS process or if there
might be differences.

\begin{acknowledgements}
The authors thank J\"orn Davidsen for helpful comments, particularly his suggestions
about the ETAS model mentioned above.
M.~B.~acknowledges the support from INFM-PAIS02. 
\end{acknowledgements}

%\bibliographystyle{egu}%<-- LIST OF REFERENCES TO BE IN "EGU" STYLE
%\bibliography{biblio_EQ2}

\end{document}